\documentclass[doublecol]{epl2}
\usepackage{graphicx}
\usepackage{color}
\usepackage{amsmath}
\usepackage{amssymb}
\usepackage{wasysym}
\pacs{62.20.mm}{}
\pacs{46.15.-x}{}
\pacs{62.20.mt}{}

\title{Crack front instabilities under mixed mode loading
in three dimensions }
\author{Herv\'e Henry\inst{1}}
\institute{\inst{1} Laboratoire Physique de la Mati\`ere Condens\'ee, \'Ecole Polytechnique, CNRS,
91128 Palaiseau, France}
\abstract{
	The evolution of a  crack front under mixed mode loading (I+III) is studied
  using a phase field model in 3 dimensions with  no stress boundary conditions. As previously observed
  experimentally in gels, there is a relaxation toward a geometry where
  $K_{III}=0$ without any front fragmentation {even} for  high values of the initial
  mode mixity $K_{III}/K_{I}$. The effects of the initial condition is studied
  and it is shown that irregularities in the initial slit can lead to front
  fragmentation for smaller values of the ratio   $K_{III}/K_{I}$ as is observed in
  experiments.
}
\begin{document}
\maketitle
\section{Introduction}

\begin{figure}
      \includegraphics[width=0.45\textwidth]{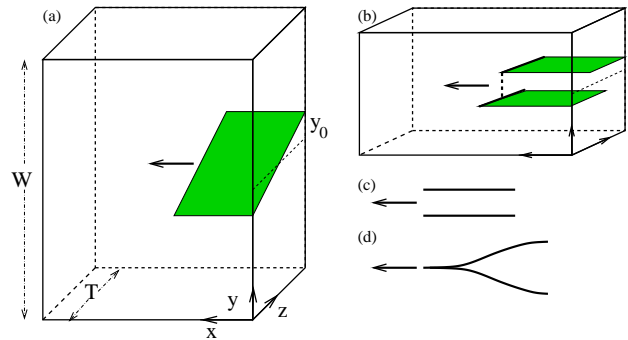}
      \caption{\label{fig_setup}\textbf{(a)}Schematic of the studied system. The initial
      crack is a plane tilted with respect to the midplane of the sample. The
      magnitude of the tilt is given by $y_0=\tan\alpha T/2$ where $\alpha$
      is the tilting angle. $W$ is the width of the sample and $T$ its
     thickness. No stress boundaries are applied at the boundary planes
     $z=\pm T/2$.  The thick  arrow
     indicates the direction of propagation of the crack as it will in other
     parts of this figure.{ \textbf{(b)} Schematic view of an echelon crack.  The two solid  thick lines
     correspond to discontinuous   crack fronts that are, here, discontinuous. The dashed
     lines are  a guide to the eye to indicate that some overlap of the fracture
     planes (and front) is possible.\textbf{(c)} is a view of the projection of
     the crack surfaces in (b) on the $xy$ plane along the z axis. And
     \textbf{(d)} is the same as (c) in the case of an \textit{\'echelon crack}
     which is closing. This case will be called a \textit{lance}.}} 
     
		\end{figure}

   While in two dimensions, the  Linear Elastic Facture Mechanics (LEFM)
   approach has been successful when describing crack propagation, its
   application to 3D systems has proven to be difficult. Indeed while in
   two dimensions, the crack front is a point and its shape cannot play any
   role in the propagation laws of the crack that relate the stress intensity
   factors to the velocity  of the crack, in three dimensions, the crack front
   shape is a new degree of freedom and its shape may affect the propagation
   law. A typical example where the crack front shape is difficult  to predict
   using the LEFM is the propagation of a crack under mixed mode (I +III).
   Indeed in this situation one expects that the cracks should eventually  propagate  in a way such that $K_{III}=0$.  Early experiments  \cite{Sommer1969} have shown
   that under mixed mode loading, continuous crack fronts can be unstable and
   may 
   become a set of discontinuous crack fronts leading to \textit{lance}
   formation (an illustration  of \textit{lances} and \textit{\'echelon} cracks
   is given in fig. \ref{fig_setup} b,c, and d). Later  theoretical \cite{pons2010,Gao1986,
   Lazarus2001,Lazarus2001b,Leblond2011,Lazarus2009}  work has shown 
   that under the
   assumption that the crack propagates in such a way that $K_{II}=0$, the
   initially flat crack front is linearly unstable in a periodic system along
   the crack front direction when the crack is under a
   load such that $K_{III}>K_{IIIc}$ where $K_{IIIc}$ is proportional to $K_I$.
   While this is  unambiguous, recent
   experiments \cite{Lazarus2001b,ronsin2014} indicate that even for high values
   of  $K_{III}/K_{I}$, smooth relaxation of the crack front toward a regime
   where  $K_{III}=0$ is possible.

    Indeed, in
    experiments by Lazarus\cite{Lazarus2001b}, where an initially  tilted crack
    crossing a finite thickness sample is
    considered (see fig. \ref{fig_setup} a) {
    there is first the initiation of
    multiple crack fronts from the  initial  crack. These fronts
    reconnect  together after having propagating over a short distance leaving
    marks that can be described as lances on the crack surface.   After their
    reconnection the continuous smooth crack front
    relaxes toward a $K_{III}=0$ configuration}. The recent experiment
    of Ronsin \cite{ronsin2014} in the same configuration showed that a   smooth relaxation
    of the crack front can  be observed without the initial \textit{lances} below a threshold tilt of the initial crack front. Above a
    threshold, crack front fragmentation occurs: an initially continuous (and
    smooth)  crack front splits into a few (at most 5) discontinuous crack front that
    propagate parallel to each other  thus forming  \textit{\'echelon cracks}.
    On the other hand, Pham
    \cite{Pham2014} observed that under mixed mode loading crack front
    fragmentation always occurs in a configuration where a linear stability
    analysis in the spirit of \cite{Gao1986} is difficult due to the complex 
    evolution of the stress intensity factors  $K_{III}$ and $K_{I}$ 
    during the propagation of the crack.  Moreover these experiments differ
    dramatically from the one of \cite{Sommer1969} since they consider finite
    thickness samples with free boundaries  and not  cylindrical samples
    that are  topologically similar to infinite periodic systems.

     Here  numerical results  on the propagation of an initially tilted crack in finite
     size samples under mode I loading are presented. They were obtained using
     the phase field model of crack
     propagation\cite{kkl,henry-04,henry-10,pons2010} applied to 3D cracks and
     show that while the phase field model is able to reproduce well the smooth
     relaxation presented in \cite{ronsin2014}, a more  detailed description of
     the experimental details is needed to actually reproduce well front
     fragmentation. Moreover they indicate that the linear instability mechanism
     presented  in\cite{pons2010,Gao1986,
   Lazarus2001,Lazarus2001b,Leblond2011,Lazarus2009} may not be at play 
    in finite size systems.{ The role of the initial crack is investigated and
    is shown to be a possible source of crack front fragmentation.}
    Before turning to the results, the system  considered and
     the phase field model are briefly described.

     \section{The phase Field model}
		The phase field approach is based on the introduction of an additional
		field $\varphi$, that describes the internal state of the material. It is
		equal to $1$ in intact regions, while it takes the value $0$ in broken
		regions. An adequate coupling between the elastic fields (the displacement)
		and the phase field is introduced so that the system has two stationary
    state solutions: one with where $\varphi=1$ and strain is uniform, and one
    where $\varphi=1$ in most of the system  except in a region of finite width (independant of the system size)
    where $\varphi$ goes smoothly  close to zero where the strain is focused\cite{kkl}.
    This is achieved through the introduction of a free energy
    functional of the form:
    \begin{equation}
      \mathcal{F}=\int \frac{D}{2}|\nabla \varphi|^2+h V_{VDW}(\varphi)+
      g(\varphi)(\mathcal {E}_{el}-\varepsilon_c^2)\label{eq_free_energy}
    \end{equation}
    Where, $g(\varphi)=4\varphi^3-3\varphi^4$, $V_{VDW}(\varphi)=\varphi^2(\varphi-1)^2$, $D$, $h$ and
    $\varepsilon_c$ are model parameters that are taken to be equal to 1 here
    (unless stated otherwise). $\mathcal {E}_{el}$ is the elastic energy density
    {in the limit of the linear elasticity}:
\begin{equation}
  \mathcal {E}_{el}=\frac{\lambda}{2}(\mathrm{tr}\varepsilon)^2+\mu
  \mathrm{tr}(\varepsilon)^2
 \end{equation}
  Here $\varepsilon$ is the stain tensor and $\mu=1$ and $\lambda=1$ the Lam\'e
 coefficients.  

    The evolution equation for the displacement field and the phase field are
    then
    \begin{eqnarray}
      \beta \dot{\varphi}&=&-max(\frac{\delta \mathcal{F}}{\delta\varphi},0) \label{evolphi}\\
      {\ddot{u_i}}&=&-\frac{\delta
      \mathcal{F}}{\delta\varphi}+D_k \Delta \dot{u}_i \label{evolui}
    \end{eqnarray}
where $\beta$ is a kinetic parameter and prescribes  the energy dissipation at
the crack front as it has been numerically shown in \cite{henry-08}. {The $D_k
\Delta \dot{u}_i$ is a phenomenological  dissipation term that has been
introduced in order to damp elastic waves  and to prevent dynamic crack
instabilities such as branching to appear without being limited to small loading
condition where lattice pinning effects are present. This term induces an
increase of the fracture energy  with its velocity as would an increase on the
value of the kinetic parameter  $\beta$\cite{henry-08}. Typical velocities of cracks are of the order of
$\approx 0.15c_s$ where $c_s$ is the shear wave speed and simulations performed
with higher load and velocities up to 0.3$c_s$ lead to the same results.}

The free parameters $D$, $h$ and $\varepsilon_c$ in
eq.\ref{eq_free_energy} allow to choose at will both the fracture energy
(denoted $\Gamma$) and the
fracture width (denoted $w_\varphi$ over which the phase field goes from 1 to
$\approx=0$).Such an approach  was initially proposed
		in \cite{aranson_crack} . Later a proper form of the coupling function
		allowing to fully relax stresses was proposed in \cite{kkl}. Since then it
		has been widely used to study the branching instability
		\cite{henry-08,karma04}, the instabilities of cracks under mixed mode in
		periodic systems \cite{pons2010}. Very similar approaches
		\cite{Francfort1998} have gained popularity in solid mechanics and
		have been used to study crack propagation. While most studies have been
		limited to two dimensional systems, the conceptual  ease of extending the
		phase field approach to three dimensional systems makes it a valuable tool
		to study crack front behaviour, a problem that has been proven to be
		difficult in the framework of LEFM. Typical example include the work of A.
    Pons on front fragmentation\cite{pons2010}, the work of Sisic on cracks induced by a
    thermal gradient and \cite{Marigo2014} where simulations have been able to
    reproduce well hexagonal columnar joints.

		Here the phase field model has been used in the geometry presented in
		figure \ref{fig_setup}. An elastic piece of material is submitted to mode I
    loading by imposing   the displacement fields in $y=\pm W/2$:
		\begin{eqnarray}
			u_x (\pm W/2) & = & 0\\
			u_y (\pm W/2) & = & \pm \Delta_y\\
			u_z (\pm W/2) & = & 0
		\end{eqnarray}
    and the mode mixity (I+III) at the crack front is due to the tilt of
    the initial crack half-plane  with respect to the mid plane of the sample
    ($y=0$). The  finite thickness of the sample imposes  the following boundary
		 conditions at $z=\pm T/2$:
  \begin{eqnarray}
		 \sigma_{zz}=0\\
     \sigma_{xz}=0\\
     \sigma_{yz}=0
     	\end{eqnarray}
     This choice of boundary condition differs dramatically  from periodic
		 boundary conditions and has been shown to affect the shape
		 of the crack front \cite{Benthem77,Bazant79,henry-10}. In samples where the
     thickness $T$  is of the same order of magnitude as its width $W$, it has been
     shown   \cite{Bazant79,henry-10} that it takes a V-Shape.

  { The boundary conditions at $x=0$ and $x=L$ are Dirichlet for the
     displacement fields and Neuman for the phase field $\varphi$ and, in order
     to study crack propagation over long distances without using too large
     simulation grids, the simulation domain was shifted by one grid point  
     order to keep the apex of the crack front close to the center of the
     simulation domain.      }
     
    { In addition into a small region of thickness $L_{ab}$  (about 40 grid points to be compared with simulation domains of length 800)
     close to the boundaries  $x=0$ and $x=L$, 
     a dissipative term  $-\alpha(x)\dot{u_i}$ term is introduced in the r.h.s. of
     eq. \ref{evolui} where $\alpha(x)$ grows linearly from zero at $x=L-L_{ab}$
     to 0.4 at $x=L$. This peculiar dissipative term was chosen to
     minimize acoustic wave reflexion at the boundaries. Simulations with
      values of $L$ multiplied by 2 show that it has no effect on the crack
      behaviour which is expected since it is only present in a region
      far from the crack front. One should note that this  also indicates that 
      the thread-mill described earlier has no effect on the crack propagation.
          
      In order to ensure that the phase field interface thickness does not
      affect  the results significantly, simulations were also performed with
      varying system size and a constant phase-field interface thickness, showing neither
      qualitative nor  quantitative change. This  indicates  a good convergence of  the model. 
      In the next section,  the results of this simulation setup in the case where the continuous
      crack front is stable and relaxes smoothly are   described. }

     \section{Results: the smooth relaxation}
 \begin{figure}
       \begin{tabular}{c}
         \includegraphics[width=0.45\textwidth]{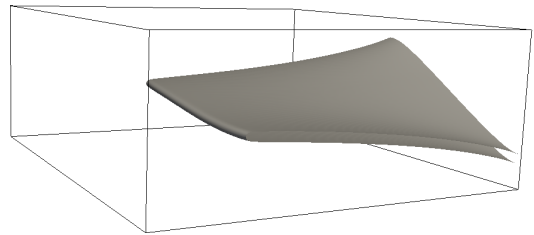}
       \end{tabular}
       \caption{\label{fig_smooth_surf} Typical crack surface for \textit{small} tilting angle
       and a smooth initial  crack surface . One
       can see that the crack front relaxes toward the mid-plane. Here all
       lengths are normalized by $T$ and the initial aspect ratio of the system
       is $0.33$. Only a portion of the simulation domain is shown. }
     \end{figure}

		 Before  turning  to the result of simulations in the case of tilted crack, recent work by Ronsin \textit{et al} in
		 nonlinear elastic gels is briefly described. In their experiments at small tilting angle (below
		 $20^o$) the crack front relaxed
		 exponentially  toward a flat crack front with a characteristic length that
		 is proportional to the size of the sample.{  Above a threshold value ranging
     from $20^o$ to $50^o$ depending on the applied load the so
     called \textit{cross hatch} instability\cite{ronsin2014} leads to crack
     front fragmentation: the propagation of two crack fronts approximately
     parallell to each other that leads to a step in the crack surface (the
     \textit{\'echelon crack}). This step eventually either drifts toward one side
     (here: $z=\pm T/2$ ) of
     the samples and vanishes or decreases leading to the reconnection of the
     two crack fronts.} Moreover, however small the initial tilting angle is, the
     crack front relaxes toward a flat crack front without any subcritical
     behaviour. 
     
\begin{figure}
       \begin{tabular}{cc}
         a & b\\
         \includegraphics[width=0.25\textwidth]{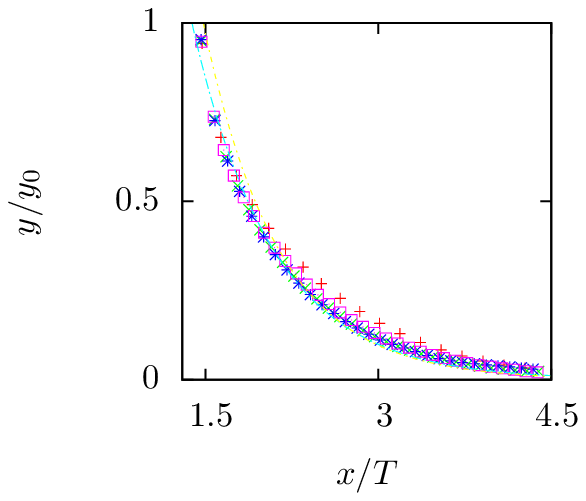}
         &
        \includegraphics[width=0.25\textwidth]{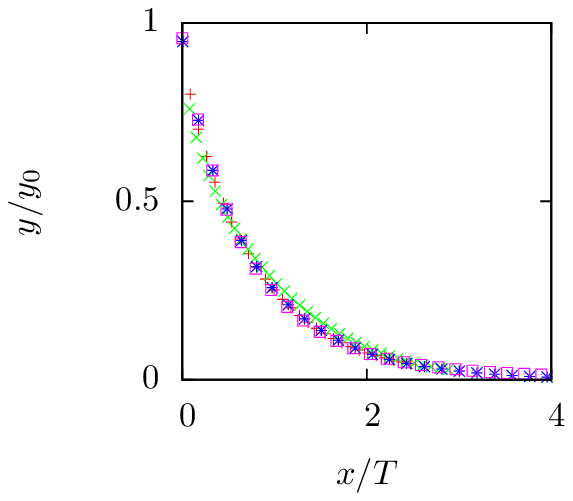}
       \end{tabular}

       \caption{\textbf{a}
       Distance between the
       intersection of the crack plane with the boundary of the elastic sample
       and the mid plane normalized by the initial distance for various
       initial tilting angles ranging from $45^o$ to $15^o$ compared with an
       exponential (used as a guide to the eye).
       There relaxation is very close to an exponential and the  characteristic
       relaxation length $l_r$ is independent
       of the initial tilt (Here $l_r\approx 0.7 T$ ). This is in very good quantitative agreement
       with results from \cite{ronsin2014}.
       \textbf{b}Same data for various sample sizes and crack velocities.
       The aspect ratios vary from 0.167 to 0.667. One can see that all curves
       collapse on a master curve.
       \label{fig_relax}}
     \end{figure}

      In good agreement with this, during the simulation performed using the
      phase field model, the smooth
     relaxation of the  crack is actually observed  as can be
     seen in figure \ref{fig_smooth_surf} without any \textit{tilt
     concentration} along the crack front.  In addition,  
    the trajectory of the intersection of the crack front with the
     sample boundary in the $xy$ plane is well fitted by a decaying  exponential.
     As in experiments these trajectories
     and  the characteristic  decay length $l_d$ are 
     independent of the initial tilt for a given sample geometry. This is
     illustrated in  fig.
     \ref{fig_relax} (a)
     where $y$, the distance from mid-plane
     has been renormalized by its initial value for different
     initial tilt angles. It can be seen that all curves collapse on a master curve.  In
     addition, the effect of changing the sample geometry (i.e. its aspect ratio
     $T/W$) was also investigated  and it appears that for aspect ratios ranging 
     from 0.167 to 0.667 the decay length is proportional to $T$ and independent
     of  $T/W$ at 
     dominant order as  it is illustrated in  fig. \ref{fig_relax} (b) where in
     addition to the normalization of $y$ by its initial value, $x$ is normalized by
     $T$, $W$ is kept constant. These results are in  good quantitative agreement with
     experimental data since the decay length measured here for an aspect ratio
     of $0.33$ is $\approx (0.65 \pm 0.05) T$ while in experiments (length was
     estimated using the value of $y$  at  $x/e=0.5$ fig. 4b in \cite{ronsin2014}) it was found
     to be equal to $(0.55\pm 0.05)T$. 

     Nevertheless, in contradiction with much experimental evidence, here an
     initially tilted  planar 
     front (below $\approx 50^o$)  never {propagated  in a way 
     leading to  a discontinuous 
     crack surface }  (independently of the system size
     $T$ or $W$).  This threshold value corresponds to an
     initial mode mixity  $K_{III}/K_{I}\approx tan \theta$ of the order unity
     while in both experiments and theoretical work the value of
     $(K_{III}/K_{I})_c$ above which an { instability } develops  is
     often much smaller (of the order 0.1).
     \begin{figure}
         \begin{tabular}{cc}
         \includegraphics[width=0.15\textwidth]{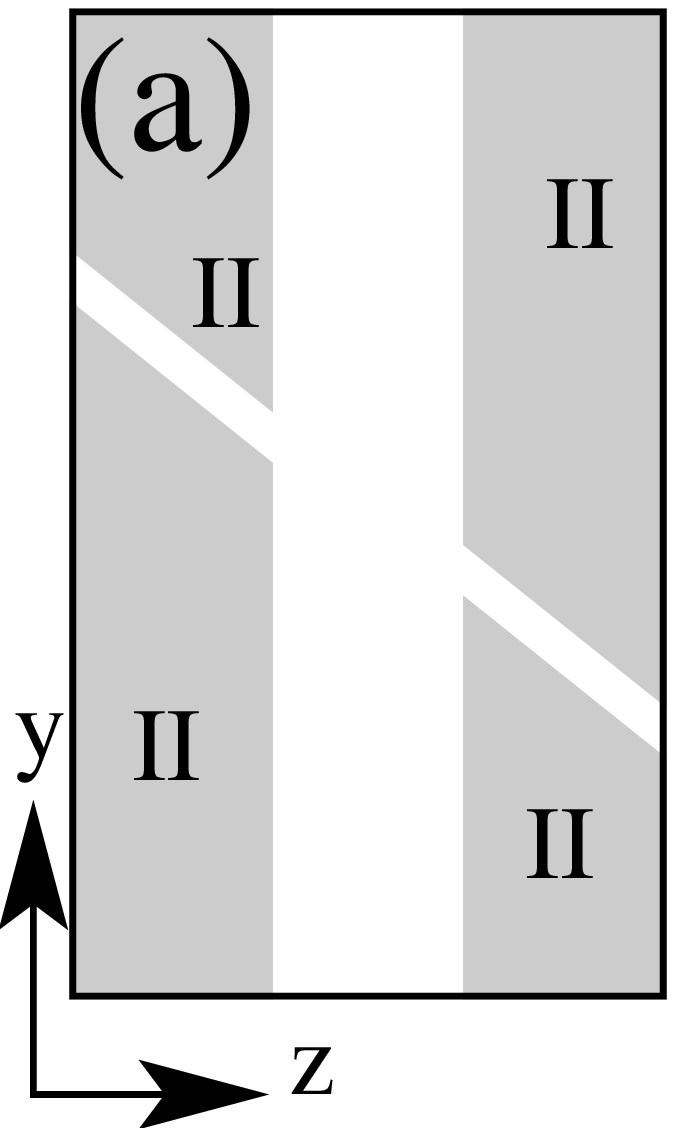}
         &
        \includegraphics[width=0.2\textwidth]{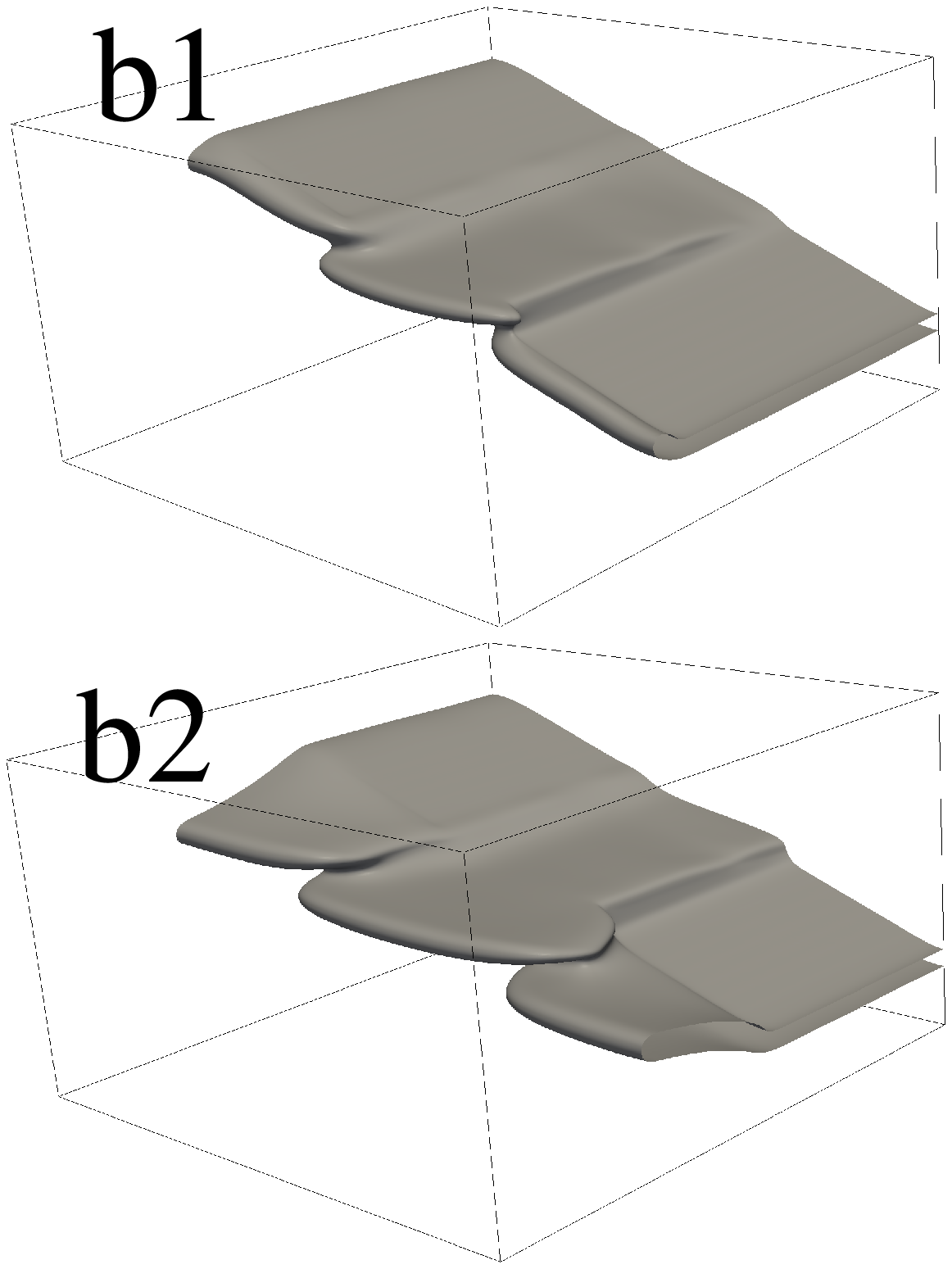}
       \end{tabular}

        \caption{\label{fig_const_tilt}(a) schematic of the fracture energy
        layout used to impose constant mode mixity. The crack is propagating
        perpendicular to the paper plane.  In
        the grey areas denoted by II, the fracture energy is 10 times higher
        than in the white region and this layout is invariant along the x axis.
        (b1) Crack surface after crack front fragmentation in this setup. (b2)
         crack surface  after propagation in a region where the fracture
        energy is again  homogeneous. In these figure, only a part of the
        simulation domain is shown. The initial crack was tilted with an angle
        of 35$^o$ and the free region extends along the z-axis over $\approx$ one
        third of the system and $\approx$ 10 times the crack interface
        thickness.}      \end{figure}

       { One should keep in mind 
       that in the simulation presented here, the
      threshold angle was independent of model parameters (keeping the fracture
      energy constant) such as the interface thickness\footnote{the interface
      thickness can be varied by changing the ratio between $D$ and both $h$ and
      $\varepsilon_c$} or the ratio
      $h/\varepsilon_c$, indicating that the discrepancy between simulation
      results and experiments is not due to a poor convergence of the phase
      field model or  to a peculiar choice of model parameters.}
      
      {In the following  a few arguments explaining 
      the discrepancy between the results presented here and
     previous theoretical and experimental  work is  given. Thereafter a  few possible
     mechanisms leading to crack front fragmentation are  investigated
     numerically. }
    
      First, the main difference between the geometry
      used here and the periodic geometry used in other theoretical or numerical  studies
      is that the  smooth relaxation
      toward a flat crack front is possible. This 
      induces an exponential decay of the mode mixity
      at the crack front. As a result, the driving force for the crack front
      linear instability decays and rapidly becomes smaller than
      the critical value above which the straight crack  front is linearly
      unstable. 
      Therefore any small perturbation of the
      initial crack front cannot grow over a sufficiently long time
      to reach a macroscopic  amplitude. In
      addition, due to boundary conditions used
      here along the z-axis (no stress instead of periodic), the crack front
      is not flat but has a curved shape (here a V-shape).  This 
      could  affect the result of a linear stability analysis and 
      may prevent the existence of the unstable modes. 
      
      In order to check this last hypothesis
     simulations were performed in a setup where the
     tilted crack front cannot relax continuously due to a simple patterning of
     the material fracture energy (see: fig. \ref{fig_const_tilt}). 
     As a result, the mode mixity is kept
     constant during propagation unless the front breaks up into  
     discontinuous lines so that
     the crack front in the central region can  relax toward an
     horizontal line. 
     { These simulations show that 
    for angles above  $\approx 20^o$ the crack front can break up
     into three  discontinuous  crack fronts if the region where the free
     propagation is possible is large enough\footnote{the critical thickness of
     the free region was found to be independent of the fracture interface
     thickness $w_\varphi$ and to depend on the imposed tilt angle} .After
     breakup, the crack front propagating into
     the \textit{free} region relaxes toward a flat surface and once the
     constraint is no longer present, the other two fronts can  relax
     toward horizontal fronts, leading to the
     propagation of three parallel crack front that form two \textit{\'echelon
     crack}.  }This shows that
     an instability mode of the crack front that can be related to the one
     depicted in \cite{pons2010,Leblond2011,Gao1986} is present in the finite
     thickness sample. It also supports the hypothesis that the absence of front
     fragmentation in simulations presented here  may be due to the fact
     that the unstable mode cannot grow sufficiently over the propagation length
     needed for the crack front to relax.  This hypothesis is confirmed   by 
     simulations where the initial crack surface was  perturbed:
     \begin{equation}
       y=\tan (\theta)z+a_0 \Sigma_n \cos(\frac{2\pi n z}{T}+\phi_n),
     \end{equation}
     where the sum is taken over 40 modes with $\phi_n$ random variables and
     $a_0$, the amplitude of the perturbation. Indeed 
     for angles as low as $38^o$   the crack front is  unstable
     for $a_0$ sufficiently high {$\approx 0.5$ (about a $1/4$  the value of $a_0$
     for which, due to discretization, the initial crack surface is discontinuous in the phase field
     model )}. This  indicates that an initial finite
     perturbation of the crack front can lead to front fragmentation.
     Nevertheless the  asymptotic behaviour of this system
     could not be investigated  due to computational  limitations. 

     In the following I present briefly an
     investigation of the effects of more dramatic alterations of the initial crack.
     This was motivated by a comparison between  experiments and the linear instability mechanism that is usually 
     proposed. {Indeed,  in the latter (except in
     \cite{ronsin2014}), the early stage is characterized by short wavelength
     facets at the onset of crack propagation that coarsen leading to
     macroscopic facets. This is unlikely to be   due to a linearly unstable mode  unless
     it has a  very short wavelength and grows over a very short distance.}.
     This leads us  to investigate other mechanisms that could lead to crack front
     fragmentation and here attention is given to the possible effects of the
     initial fracture.
     
         \begin{figure}
    \includegraphics[width=0.4\textwidth]{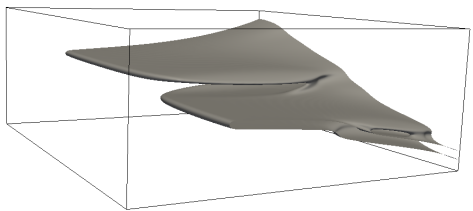}
    \caption{\label{fig_drift} Crack surface observed after the crack front that
    was initially tilted and perturbed has
    broken into two well defined crack fronts. The size of the box is the same
    as the  the one of fig. \ref{fig_smooth_surf}, where the initial tilt had
    almost completely relaxed.
    The initial tilting angle was $45^o$ which is  close to the
    threshold angle for front fragmentation which is about $50^o$. Similar
    patterns were observed with lower initial tilt angle and higher noise.}
  \end{figure}
  \section{Effect of the initial condition on the fragmentation process}
\begin{figure}
        \centerline{ \includegraphics[width=0.5\textwidth]{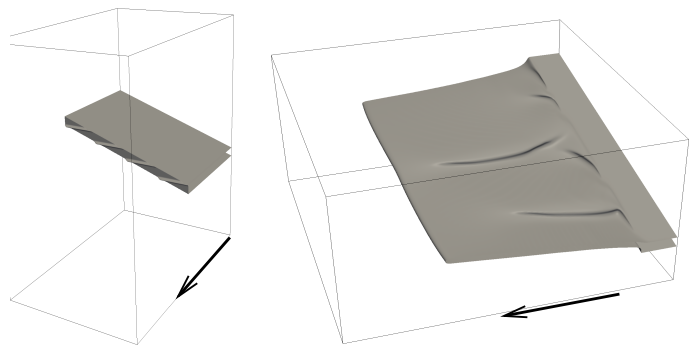}
}
\caption{\label{fig_ini_slit}\textbf{Left} Example of initial $\varphi=0.5$
iso-surface that were used to study the
 effect of the initial slit shape on the crack front fragmentation
 process.\textbf{Right} Crack surface after propagation initiated from such an
 initial condition. The  tilt angle is 25$^o$. }
\end{figure}

     In all simulations presented previously the initial crack is rounded
    at its tip so that the initial crack front is well defined and continuous. 
    Here we present simulations where the initial cracked region   has a  flat
    front  with some protrusions in
    the direction of the crack propagation as illustrated in fig.
    \ref{fig_ini_slit}-a. As a result  many disconnected crack
    fronts can begin to propagate, leading possibly to discontinuous crack
    fronts propagation and eventually coarsening as reported for instance  in\cite{Karma2015}. 

 \begin{figure}
    \includegraphics[width=0.4\textwidth]{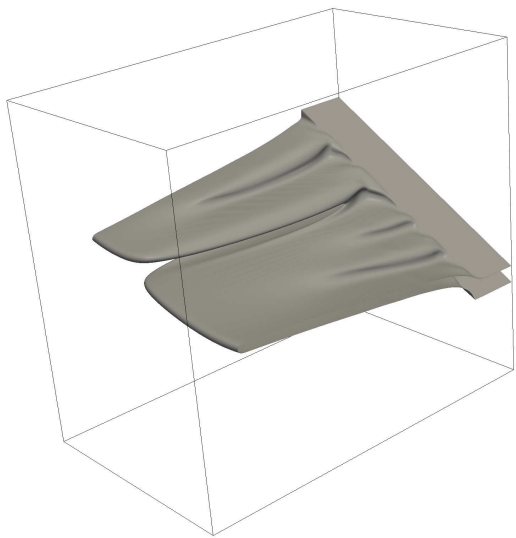}
    \caption{\label{fig_mess} Crack surface after the crack has propagated over
    approximately twice the sample thickness initiated by an initial slit with
     evenly distributed asperities. The evolution of the system leads to the
     propagation of two  well defined crack fronts and the trace of other crack
     front discontinuities that have  smeared out  are  still well marked. The
     initial tilt angle was 35$^o$.}
  \end{figure}

     For  this purpose we consider
     initial slits whith thickness of roughly 5 times $w_\varphi$
     and  at the surface of which a few irregularities are present: lines of
     crack with thickness and length $w_\varphi$\footnote{flat and rounded  initial slits,
     were also investigated, showing no significant difference from the approach
     described earlier}. These simulations mimic  the propagation of a crack
     front initiated by an initial slit machined
     in a solid with some imperfections and can also be seen as a caricature of
     a crack propagating in a medium where  structural heterogeneities  
     are present\cite{ronsin2014}. Due to computational limitations, these
     simulations have been restricted  to systems where a small  number of  defects
     (at most 20) of the front are present. In the following the results are
     briefly described depending on the initial configuration. The simplest case
     considered is an initial front with  two
     \textit{horizontal} (i.e. $y=$ constant) defects. The evolution
     of the crack front leads, after a  short
     transient, to the growth  of two well defined crack surfaces.
      In the
     situation where many ($\approx 10$ ) initial cracks are present, the
     evolution of the crack front leads to the propagation of multiple, close to 
     horizontal, crack surfaces that propagate forming well defined
     \textit{\'echelons cracks}. During their propagation  coarsening
     (here coalescence of two crack fronts and not elimination of a crack
     surface) leads to the propagation of a diminishing number of
     crack fronts  that are separated by significantly higher gap) 
     leading to patterns  very similar to the one described in
     \cite{Sommer1969,Karma2015,pons2010}. 
     Nevertheless the setup and the hypothesis here 
     are extremely different:   the boundary conditions are no stress instead
     of periodic and  the initial crack surface is \textit{significantly
     perturbed} with asperities that act as \textit{nuclei} for the crack front
     and not flat.  This implies that  the linear instability scenario is not
     the sole explanation for crack front fragmentation and that another
     possibility is the nucleation of multiple crack fronts either due to
     irregularities in  the initially machined slit or {to material
     inhomogeneities (this possibility was not investigated here) but has been
     shown to be very likely in \cite{ronsin2014} where the effect of the
     initial crack was carefully investigated and shown not to be significant.}

     \section{Conclusion}
      In summary the propagation of a crack front under mixed mode loading in a
      finite sample with no stress boundary condition has been investigated. The
      results indicate that smooth relaxation of the initial crack front is
      observed below a threshold mode mixity (here tilt  angle of the initial
      slit)  that is  below
      the one observed experimentally\footnote{Ronsin \textit{et al} indicate
      that they have observed examples of smooth relaxation up to angles of $50^o$}
      unless the initial crack front is significantly perturbed.  In this latter
      case, the crack propagation patterns are similar to those observed
      experimentally under various conditions where no stress boundary
      conditions are present. This indicates that the linear instability
      mechanism initially  presented in \cite{Gao1986} and further
      refined\cite{Lazarus2001,Lazarus2001b,pons2010,Leblond2011}
      may not apply to situations where boundary conditions do not prevent the
      smooth relaxation of the front as observed experimentally. 
      Moreover simulations where the initial slit is not \textit{uniform} give a
      possible explanation for the front fragmentation observed in some
      experiments where the material is  inhomogeneous (e.g. cavities already
      present in the material
      can act as perturbation of the crack front)  or where the initial
      slit is inhomogeneous or thick compared to the process zone.
      { This
      calls for  experiments where both the defects of the 
      initial slit and the structural
      inhomogeneities are controlled and have characteristic length 
      scales that can be tuned as in\cite{VDBcrack}.}

\acknowledgments
 I wish to thank Allistair Rowe and Mathis Plapp for their useful comments on
 early version of the manuscript. I  wish to thank O. Ronsin, T. Baumberger
 and C. Caroli for enlighting  discussions on their experimental work at an
 early stage of this work.

\bibliographystyle{eplbib}

		 \end{document}